\documentclass[preprint,showpacs,showkeys,preprintnumbers,aps]{revtex4}

\usepackage{graphicx}

\begin{document}

\preprint{Balke et al., Co$_2$Mn$_{1-x}$Fe$_x$Si.}

\title{Properties of the quaternary half-metal-type Heusler alloy Co$_2$Mn$_{1-x}$Fe$_x$Si.}

\author{Benjamin Balke, Gerhard H. Fecher, Hem C. Kandpal, and Claudia Felser}
\email{fecher@uni-mainz.de}
\affiliation{Institut f\"ur Anorganische und Analytische Chemie, \\
             Johannes Gutenberg - Universit\"at, D-55099 Mainz, Germany.}

\author{Keisuke Kobayashi, Eiji Ikenaga, Jung-Jin Kim, and Shigenori Ueda}
\affiliation{Japan Synchrotron Radiation Research Institute (SPring-8/JASRI), \\
             Kouto 1-1-1, Mikaduki-cho, Sayou-gun, Hyogo, 679-5198, Japan}
\date{\today}

\begin{abstract}

This work reports on the bulk properties of the quaternary Heusler
alloy Co$_2$Mn$_{1-x}$Fe$_x$Si with the Fe concentration $x=$. All
samples, which were prepared by arc melting, exhibit $L2_1$ long range
order over the complete range of Fe concentration. Structural and
magnetic properties of Co$_2$Mn$_{1-x}$Fe$_x$Si Heusler alloys were
investigated by means of X-ray diffraction, high and low temperature
magnetometry, M{\"o\ss}bauer spectroscopy, and differential scanning
calorimetry. The electronic structure was explored by means of high
energy photo emission spectroscopy at about 8~keV photon energy. This
ensures true bulk sensitivity of the measurements. The magnetization of
the Fe doped Heusler alloys is in agreement with the values of the
magnetic moments expected for a Slater-Pauling like behavior of
half-metallic ferromagnets. The experimental findings are discussed on
the hand of self-consistent calculations of the electronic and magnetic
structure. To achieve good agreement with experiment, the calculations
indicate that on-site electron-electron correlation must be taken into
account, even at low Fe concentration. The present investigation
focuses on searching for the quaternary compound where the
half-metallic behavior is stable against outside influences. Overall,
the results suggest that the best candidate may be found at an iron
concentration of about 50\%.

\end{abstract}



\maketitle

\section{Introduction}

K{\"u}bler {\it et al} \cite{KWS83} recognized that the minority-spin
state densities at the Fermi energy  nearly vanish for Co$_2$MnAl and
Co$_2$MnSn. The authors concluded that this should lead to peculiar
transport properties in these Heusler compounds because only the
majority density contributes. The so called half-metallic ferromagnets
have been proposed as ideal candidates for spin injection devices
because they have been predicted to exhibit 100~{\%} spin polarization
at the Fermi energy ($\epsilon_F$) \cite{GME83}. From the applications
point of view, a high Curie temperature for a half-metallic ferromagnet
may be an important condition. For this reason, Heusler alloys with
$L2_1$ structure have attracted great interest. Some of these alloys
exhibit high Curie temperatures and, according to theory, should have a
high spin polarization at the Fermi energy \cite{GDP02,PCF02,FKW06}.
Calculations also show that anti-site disorder will destroy the high
spin polarization \cite{PCF04a}, implying that precise control of the
atomic structure of the Heusler alloys is required.

The Heusler alloy Co$_2$MnSi has attracted particular interest because
it is predicted to have a large minority spin band gap of 0.4~eV and,
at 985~K, has one of the highest Curie temperature,  among the known
Heusler compounds \cite{FSI90,BNW00}. Structural and magnetic
properties of Co$_2$MnSi have been reported for films and single
crystals \cite{RRH02,GBW02,KHM03,SBM04b,WPK05a,WPK05b}. In accordance
with theoretical predictions, bulk Co$_2$MnSi has been stabilized in
the $L2_1$ structure with a magnetization of 5~$\mu_{B}$ per formula
unit. From tunneling magneto resistance (TMR) data with one electrode
consisting of a Co$_2$MnSi film Schmalhorst {\it et al}
\cite{SKS04,SKR05} inferred a spin polarization of 61~{\%} at the
barrier interface. Although the desired spin polarization of 100~{\%}
was not reached, the experimental value of the spin polarization is
larger than the maximum 55~{\%} effective spin polarization of a
variety of 3$d$-transition metal alloys in combination with Al$_2$O$_3$
barriers \cite{LSK00}. However, the spin polarization of photoelectrons
emerging from single crystalline Co$_2$MnSi films grown on GaAs by
pulsed laser deposition indicate a quite low spin polarization at the
Fermi level of only 12~{\%} at the free surface \cite{WPK05b}. Wang
{\it et al} \cite{WPK05a,WPK05b} assumed that partial chemical disorder
was responsible for this discrepancy with the theoretical predictions.

Photo emission spectroscopy is the method of choice to study the
occupied electronic structure of materials. Low kinetic energies result
in a low electron mean free path being only 5.2~\AA at kinetic energies
of 100~eV (all values calculated for Co$_2$FeSi using the TPP2M
equations \cite{TPP93}) and a depth of less than one cubic Heusler cell
will contribute to the observed intensity. The situation becomes much
better at high energies. In the hard X-ray region of about 8~keV one
will reach a high bulk sensitivity with an escape depth being larger
than 115~\AA (corresponding to 20 cubic cells). High energy photo
emission (at about 15~keV excitation energy) was first performed as
early as 1989 \cite{Mei89} using a $^{57}$Co M\"o\ss bauer
$\gamma$-source for excitation, however, with very low resolution only.
Nowadays, high energy excitation and analysis of the electrons become
easily feasible due to the development of high intense sources
(insertion devices at synchrotron facilities) and multi-channel
electron detection. Thus, high resolution - high energy photo emission
(HRHEPE) was recently introduced by several groups
\cite{KYT03,SS04,TKC04,Kob05,PCC05,TSC05} as a bulk sensitive probe of
the electronic structure in complex materials. In the present work,
HRHEPE at $h\nu\approx8$~keV was used to study the density of states of
Co$_2$Mn$_{1-x}$Fe$_{x}$Si with $x=0,1/2,1$.

Recent investigations \cite{WFK05,WFK06a,WFK06b,KFF06} of the
electronic structure of Heusler compounds indicate that on-site
correlation plays an important role in these compounds and may serve to
destroy the half-metallic properties of Co$_2$MnSi. In addition, if
on-site correlation is considered in electronic structure calculations
Co$_2$FeSi becomes a half-metallic ferromagnet with a magnetic moment
of 6~$\mu_{B}$.

The present investigation focuses on searching for a mixed compound in
the series Co$_2$Mn$_{1-x}$Fe$_x$Si where the half-metallic behavior is
stable against the variation of on-site correlation and other outside
influences.

\section{Computational Details}
\label{sec:CD}

The present work reports, besides experiments, on calculations of the
electronic and magnetic properties of ordered Heusler compounds of the
Co$_2$(Mn$_{(1-x)}$Fe$_{x}$)Si type. The random alloys were treated as
virtual crystals of the Co$_2$Mn$_{(1-i/4)}$Fe$_{i/4}$Si type with
$i=0,1,2,3,4$. Non-rational values of $x$ as well as random disorder
(for examples see references \cite{MNS04,KUK04,FKW05}) will not be
discussed here.

The self-consistent electronic structure calculations were carried out
using the scalar-relativistic full potential linearized augmented plane
wave method (FLAPW) as provided by Wien2k \cite{BSM01}. In the
parameterization of Perdew {\it et al} \cite{PCV92} the
exchange-correlation functional was taken within the generalized
gradient approximation (GGA). A $20\times20\times20$ mesh was used for
integration of cubic systems, resulting in 255 $k$-points in the
irreducible wedge of the Brillouin zone.

The properties of pure compounds containing Mn or Fe were calculated in
$F\:m\overline{3}m$ symmetry using the experimental lattice parameter
($a=10.658 a_{0B}$, $a_{0B} = 0.529177$~{\AA}) determined by X-ray
powder diffraction. Co atoms are placed on 8c Wyckoff positions, Mn or
Fe on 4a and Si on 4b \footnote{Note that 4a and 4b positions are
equivalent; for clarity we assume that Si is always on 4b.}. All muffin
tin radii were set as nearly touching spheres with $r_{MT}=2.3 a_{0B}$.
A structural optimization for the pure compounds showed that the
calculated lattice parameter deviates from the experimental one only
marginally.

The calculation of mixed random alloys is not straight forward in the
FLAPW as is used here. However, the substitution of some Mn atoms of
the $L2_1$ structure by Fe leads in certain cases to ordered structures
that can be easily used for the calculations. Those ordered, mixed
compounds have the general formula sum
Co$_8$(Mn$_{(1-x)}$Fe$_x$)$_4$Si$_4$ and have integer occupation of Mn
and Fe if $x=i/4$ where $i=1,2,3$. (for more details see
Ref.\cite{FKW05}). To verify that ordered compounds could be used
instead of random alloys, the full relativistic Korringa-Kohn-Rostocker
(KKR) method with the coherent potential approximation (CPA) was
employed \cite{Ebe99}. The exchange-correlation functional was
parameterized by using the plain GGA. No significant differences in the
integrated properties, such as the density of states or the magnetic
moments, were found between the methods.

For the case of Co$_2$FeSi, it has recently been demonstrated that LSDA
or GGA schemes are not sufficient for describing the electronic
structure correctly. Significant improvement was found, however, when
the LDA$+U$ method \cite{WFK05, KFF06} was used and this computational
scheme was used here as well. LDA$+U$, as described by Anisimov {\it et
al} \cite{AAL97}, adds an orbital dependent electron-electron
correlation, which is not included in the plain LSDA or GGA schemes. It
should be mentioned that the $+U$ was used in the FLAPW scheme with the
GGA rather than the LSDA parameterization of the exchange-correlation
functional. No significant differences were observed using either of
these parameterizations.

\section{Experimental Details}
\label{sec:ED}

Co$_2$Mn$_{1-x}$Fe$_x$Si samples were prepared by arc melting of
stoichiometric amounts of the constituents in an argon atmosphere at
10$^{-4}$~mbar. Care was taken to avoid oxygen contamination. This was
ensured by evaporating Ti inside of the vacuum chamber before melting
the compound as well as by additional purifying of the process gas. The
polycrystalline ingots that were formed were then annealed in an
evacuated quartz tube for 21~days. This procedure resulted in samples
exhibiting the Heusler type $L2_1$ structure, which was verified by
X-ray powder diffraction (XRD) using excitation by Cu K$_\alpha$ or Mo
K$_\alpha$ radiation.

Flat disks were cut from the ingots and polished for spectroscopic
investigations of bulk samples. For powder investigations, the
remainder was crushed by hand using a mortar. It should be noted that
using a steel ball mill results in a strong perturbation of the
crystalline structure.

X-ray photo emission (ESCA) was used to verify the composition and to
check the cleanliness of the samples. After removal of the native oxide
from the polished surfaces by Ar$^+$ ion bombardment, no impurities
were detected with ESCA. The samples were afterwards capped in-situ by
a 2~nm layer of Au at room temperature to prevent oxidation of the
samples during transport in air.

Magneto-structural investigations were carried out using M{\"o\ss}bauer
spectroscopy in transmission geometry using a constant acceleration
spectrometer. For excitation, a $^{57}$Co(Rh) source with a line width
of 0.105~mm/s (5~neV) was used. The spectra from powder samples were
taken at 290~K.

The magnetic properties were investigated by a super conducting quantum
interference device (SQUID, Quantum Design MPMS-XL-5) using nearly
punctual pieces of approximately 5~mg to 10~mg of the sample.
Differential scanning calorimetry (DSC) measurements (NETZCH, STA 429)
were performed to detect phase transitions below the melting point. In
particular, attempts were made to find the Curie temperature ($T_C$),
but this turned out to be too high to be determined directly by the
SQUID, which is limited to 775~K even in the high temperature mode.

The electronic structure was explored by means of high energy X-ray
photo emission spectroscopy. The measurements were performed at the
beamline BL47XU of the synchrotron SPring~8 (Hyogo, Japan). The photons
are produced by means of a 140-pole in vacuum undulator and are further
monochromized by a double double-crystal monochromator. The first
monochromator uses Si(111) crystals and the second a Si(111)
channel-cut crystal with 444 reflections (for 8~keV X-rays). The energy
of the photo emitted electrons is analyzed using a Gammadata - Scienta
R~4000-12kV electron spectrometer. The ultimate resolution of the set
up (monochromator plus analyzer at 50~eV pass energy using a 200~$\mu$m
slit) is 83.5~meV at 7935.099~eV photon energy. For the here reported
experiments, a photon energy of 7939.15~eV  has been employed. Under
the present experimental conditions an overall resolution of 250~meV
has been reached. All values concerning the resolution are determined
from the Fermi-edge of an Au sample. Due to the low cross-section of
the valence states from the investigated compounds, the spectra had to
be taken with $E_{pass}=200$~eV and a 500~$\mu$m slit for a good signal
to noise ratio. The polycrystalline samples have been fractured in-situ
before taking the spectra to remove the native oxide layer. Core-level
spectra have been taken to check the cleanliness of the samples. No
traces of impurities were found. The valence band spectra shown in
section \ref{sec:EP} were collected over 2-4~h at about 100~mA electron
current in the storage ring in the top-up mode. All measurements have
been taken at a sample temperature of 20~K.

\section{Results and Discussion}
\subsection{Electronic and Magnetic Structure}
\label{sec:EMS}

The electronic structure of the substitutional series
Co$_2$Mn$_{(1-i/4)}$Fe$_{i/4}$Si with $i=0,\ldots,4$ was calculated
using the LDA$+U$ method. This method was used because it was found
that plain GGA calculations are not sufficient to explain the magnetic
moments in Co$_2$FeSi \cite{WFK05}. Using the impurity model of
Anisimov and Gunnarsson \cite{AGu91} along with self consistent
calculations, the effective Coulomb-exchange parameter $U_{eff}$ was
determined for the pure compounds containing Mn and Fe. Details of the
procedure and the implementation of the constrained LDA calculations in
FLAPW are reported by Madsen and Novak \cite{MNo05}. The values found
in the present work are $U_{Co}=0.30$~Ry and $U_{Mn}=0.39$~Ry for
Co$_2$MnSi, and $U_{Co}=0.31$~Ry and $U_{Fe}=0.32$~Ry for Co$_2$FeSi.

Comparing the semi-empirical values used in \cite{KFF06} to the values
found here from the constrained LDA calculations, it is evident that
the latter are too high to explain the magnetic moments. Additional
calculations for the elemental $3d$ transition metals revealed that all
values for $U_{eff}$ found in the constrained LDA calculations are
considerably too high to explain those metallic systems correctly. This
is despite the fact that such calculations may result in reliable
values for Mott insulators \cite{AZA91}.

For these reasons, the semi-empirical values corresponding to 7.5~\% of
the atomic values (see: Ref.\cite{KFF06}) will be used and discussed
here for the case of the FLAPW calculations. These values ensure that
the calculated magnetic moments agree with the measured  values over
the entire range of the Fe concentration (compare Sec.:~\ref{sec:MP}).
In particular, the values for $U_{eff}$ were set to $U_{Co}=0.14$~Ry,
$U_{Fe}=0.132$~Ry, and $U_{Mn}=0.13$~Ry, independent of the iron
concentration. These values are closer to the values for the Coulomb
interaction $U_{dd}$ for $d$ electrons in the elemental $3d$ transition
metals reported by Bandyopadhyay and Sarma \cite{BSa89} even before the
LDA$+U$ method itself was introduced.

The use of $U_eff=U-J$ suppresses multipole effects. That means, it
neglects the non-spherical terms in the expansion of the Coulomb
interaction. Additionally, full potential linearized muffin tin
orbitals (FPLMTO) calculations were performed to check for the
influence of the non-spherical terms. Differently, the LMTART 6.5
program provided by Savrasov \cite{Sav96} uses the Slater integrals
$F^0 \ldots F^4$ for the calculation of the on-site correlation with
$U=F^0$ and $J=(F^2+F^4)/14$ (see also Refs.:\cite{AGu91,AAL97}).

As in FLAPW, the use of the $U$ values from the constrained LDA
calculations leads also in FPLMTO to much too large values for the
magnetic moments compared to the experimental findings. In a next step,
only reduced values for $F^0$ were used. Following the arguments of
Ref.:\cite{ASK93}, only $F^0$ should be effected by the screening in
the solid state. However, the use of the atomic values for $F^2$ and
$F^4$ as proposed in Ref.:\cite{ASK93} still did not lead to
satisfactory values for the magnetic moments. Finally, the reduction of
all Slater integrals to 10~\% of their atomic values was leading to
results being compatible to the measured magnetic moments. Moreover,
nearly identical results as in FLAPW were obtained concerning not only
the magnetic moments but also the band structures. The slightly higher
values resulting in the most appropriate for FPLMTO are caused by the
fact that these calculations used the $U$ on top of the LDA
exchange-correlation functional and not on top of the GGA
parameterization. This behavior reveals that GGA includes some more
correlation compared to pure LDA. It is also observed within the FLAPW
scheme.

The different values for $U$, $U_{dd}$, $U_{eff}$, and $F^0 \ldots F^4$
are summarized in Table \ref{tab00}. The values according to
Ref.:\cite{BSa89} where calculated for the $d$-state occupation as
found from calculations using spherical potentials. The values for
FLAPW are calculated from the Slater integrals reduced to 7.5\% of the
atomic values as calculated by Cowans program \cite{Cow81, KFF06} and
the values for FPLMTO correspond to 10\% of the atomic values. The use
of these values leads to nearly identical results comparing FLAPW and
FPLMTO.

\begin{table*}
\centering
\caption{$LDA+U$ parameter for Co$_2$Mn$_{1-x}$Fe$_x$Si. \\
         (All values are given in Ry.)}

     \begin{tabular}{l|c|c|c|ccc}
                & Ref.:\cite{BSa89} & constrained LDA & FLAPW     & FPLMTO &       &       \\
        element & $U_{dd}$          & $U$             & $U_{eff}$ & $F^0$  & $F^2$ & $F^4$ \\
        \noalign{\smallskip}\hline\noalign{\smallskip}
        Co      &  0.194            & 0.3(1)          & 0.140     & 0.185  & 0.085 & 0.053 \\
        Fe      &  0.171            & 0.32            & 0.132     & 0.175  & 0.081 & 0.050 \\
        Mn      &  0.148            & 0.39            & 0.130     & 0.164  & 0.076 & 0.047 \\
        \noalign{\smallskip}\hline
    \end{tabular}
    \label{tab00}
\end{table*}

\begin{figure*}
\centering
\includegraphics[width=15.5cm]{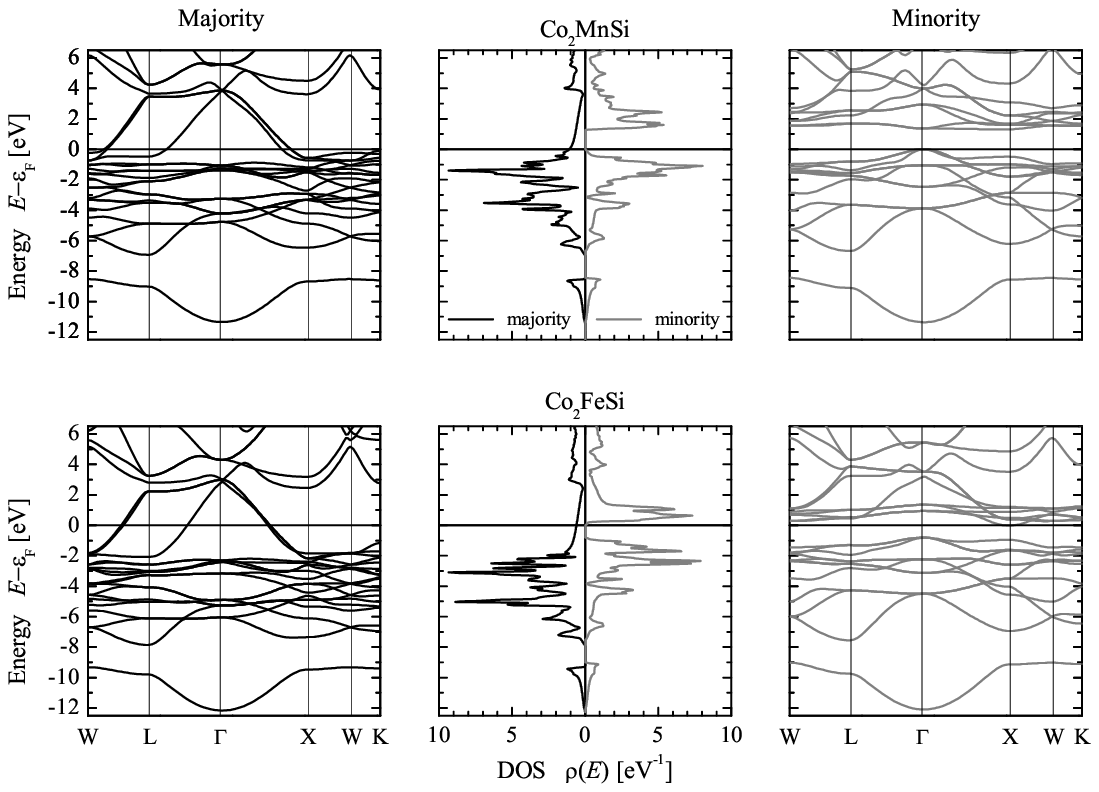}
\caption{Band structure and density of states of Co$_2$MnSi and
         Co$_2$FeSi.}
\label{fig1}
\end{figure*}

In the following, only the results from the FLAPW calculations are
discussed to allow for a better comparison with previous work. However,
all results shown below are compatible and agree very well with those
using LDA$+U$ in FPLMTO calculations. Figure \ref{fig1} shows the spin
resolved band structure and the total density of states for the pure
compounds Co$_2$MnSi and Co$_2$FeSi as calculated within the framework
of the LDA$+U$. In all cases, the band structures are very similar and
the gap in the minority bands is clearly revealed.

When explaining the Heusler half-metallic ferromagnets using simple
rigid band-like or molecular orbital-like models, it is expected that
the additional $d$ electron of the Fe compound fills the majority
states while not affecting the minority states. As may be seen from
Fig.:~\ref{fig1}, this is clearly a strong oversimplification. The
additional electron must be absorbed in the strongly dispersing
unoccupied $d$-bands seen in the Mn compound just above $\epsilon_F$.
Comparing the majority DOS, it can be seen that the high density
$d$-states at -1.4~eV (or -3.5~eV) in Co$_2$MnSi are shifted to
approximately -3~eV (or -5~eV) in Co$_2$FeSi. Keeping the minority DOS
fixed, this would imply an additional exchange splitting of about
1.6~eV, when compared to Co$_2$MnSi, between the majority and minority
states in Co$_2$FeSi. This large and rather unphysical shift indicates
that the rigid band model fails and that other alterations of the band
structure must take place.

After inspecting the electronic structure in more detail, some
particular changes are found. For example, the rather large shift of
the occupied majority $d$-states is compensated by a shift of the
occupied minority $d$-states, and this keeps the exchange splitting
rather fixed. This then results in a shift and splitting of the
occupied minority $d$-states (seen at -1~eV in the Mn compound and at
-1.7~eV or -2.3~eV in the Fe compound) as well as a shift of the
unoccupied minority $d$-states towards the Fermi energy. In addition,
the splitting of the unoccupied minority $d$-states just above the gap
is reduced from 0.7~eV in Co$_2$MnSi to 0.4~eV in Co$_2$FeSi. The most
striking effect, however, is the shift of the Fermi energy from the top
of the minority valence band to the bottom of the minority conduction
band. These particular positions of the minority gap with respect to
the Fermi energy make both systems rather unstable with respect to
their electronic and magnetic properties. Any small change of a
physically relevant quantity may serve to destroy the HMF character by
shifting the Fermi energy completely outside of the minority gap. As
long as the shift is assumed to be small, the magnetic moment may still
be similar to the one expected from a Slater-Pauling behavior, even so,
the minority gap is destroyed. For this reason, the magnetic moment may
not provide evidence for a half-metallic state.

\begin{figure}
\centering
\includegraphics[width=7.5cm]{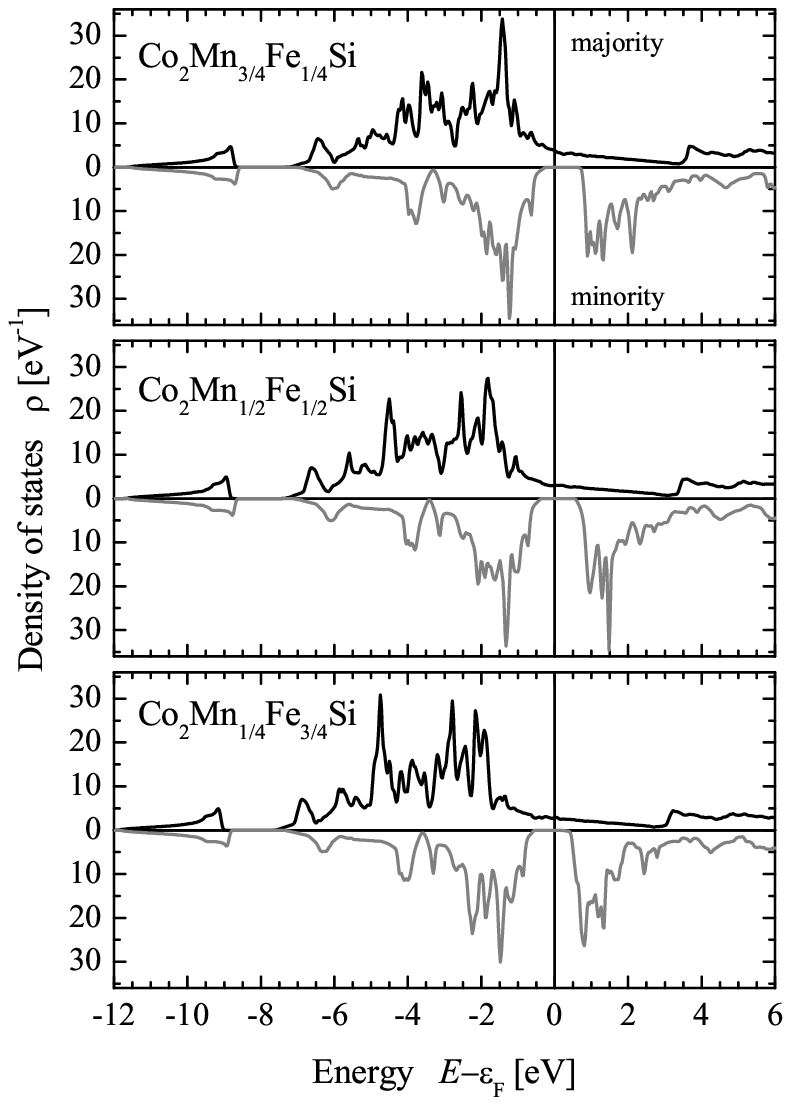}
\caption{Spin  resolved density of states of
         Co$_2$Mn$_{1-x}$Fe$_{x}$Si for $x=1/4,1/2,$ and $3/4$.}
\label{fig2}
\end{figure}

It is to be immediately expected that the situation improves in the
mixed compounds containing both Mn and Fe. Figure \ref{fig2} shows the
spin resolved total density of states for the compounds with an
intermediate Fe concentration ($x\neq0,1$). In all cases, the gap in
the minority bands is kept.

The shift of the majority $d$-states with low dispersion away from the
Fermi energy is clearly visible in Fig:~\ref{fig2}. The additional
charge (with increasing Fe concentration $x$) is filling the strongly
dispersing $d$-states in the majority channel. At the same time, the
minority DOS is shifted with respect to the Fermi energy such that
$\epsilon_F$ moves from the top of the minority valence bands at low
$x$ to the bottom of the minority conduction bands at high $x$. In
general, it can be concluded that the additional electrons affect both
majority and minority states.

\begin{table}
\centering
\caption{Properties of the minority gap of ordered Co$_2$Mn$_{1-x}$Fe$_x$Si. \\
         Given are the extremal energies of the valence band maximum (VB$_{max}$),
         the conduction band minimum (CB$_{min}$), and the resulting gap ($\Delta E$)
         in the minority states as found from LDA$+U$ calculations.
         The extremal energies are given with respect to $\epsilon_F$. All energies are given in eV.}

     \begin{tabular}{l|ccc}
        $x$   & VB$_{max}$  & CB$_{min}$ & $\Delta E$ \\
        \noalign{\smallskip}\hline\noalign{\smallskip}
        0     &   0.003     & 1.307      & 1.3  \\
        $1/4$ &  -0.181     & 0.970      & 1.15 \\
        $1/2$ &  -0.386     & 0.495      & 0.88 \\
        $3/4$ &  -0.582     & 0.181      & 0.86 \\
        1     &  -0.810     &-0.028      & 0.78 \\
        \noalign{\smallskip}\hline
    \end{tabular}
    \label{tab0}
\end{table}

Table \ref{tab0} summarizes the results for the gap in the minority
states as found from LDA$+U$ calculations. The largest gap in the
minority states is found for Co$_2$MnSi. The size of the gap decreases
with increasing Fe content $x$, and at the same time, the position of
the Fermi energy is moved from the top of the valence band to the
bottom of the conduction band. It is also seen that the compounds with
$x=0$ and 1 are on the borderline to half-metallic ferromagnetism, as
the Fermi energy just touches the top of the valence band or the bottom
of the conduction band. In both cases, a slight change of $U_{eff}$ in
the calculation is able to shift $\epsilon_F$ outside of the gap in the
minority states.

For intermediate Fe concentration, the Fermi energy falls close to the
middle of the gap in the minority states (see also Fig. \ref{fig2}).
This situation makes the magnetic and electronic properties of the
compound very stable against external influences that will not be able
to change the number of minority electrons. This applies both to the
parameters in the theoretical calculations as well as the actual
experimental situation. From this observation it can be concluded that
Co$_2$Mn$_{1/2}$Fe$_{1/2}$Si exhibits a very stable half-metallic
character in this series of compounds, as well as those with a
concentration close to $x=0.5$.

If the Heusler alloys are half-metallic ferromagnets, then they will
show a Slater-Pauling behavior for the magnetization, meaning that the
saturation magnetization scales with the number of valence electrons
\cite{GDP02,FKW06,JKW00}. The magnetic moment per unit cell (in
multiples of the Bohr magneton $\mu_B$) is given by:

\begin{equation}
   m = N_V - 24,
\label{eq1}
\end{equation}

with $N_V$ denoting the accumulated number of valence electrons in the
unit cell. For Co$_2$MnSi there is a total of $2\times9+7+4=29$ valence
electrons in the unit cell and accordingly 30 for Co$_2$FeSi. for this
reason the magnetic moment is expected to vary linearly from 5~$\mu_B$
to 6~$\mu_B$ with increasing iron concentration in
Co$_2$Mn$_{1-x}$Fe$_x$Si.

\begin{table}
\centering
\caption{Total magnetic moments of ordered Co$_2$Mn$_{1-x}$Fe$_x$Si. \\
         All moments were calculated for the given super-cells.
         Their values are in $\mu_B$ and respect 4 atoms in the unit cell for easier comparison.}
     \begin{tabular}{l|c|cc}
        compound             & $x$ & GGA  & LDA$+U$ \\
        \noalign{\smallskip}\hline\noalign{\smallskip}
        Co$_2$MnSi           & 0     & 5.00 & 5.00 \\
        Co$_8$Mn$_3$FeSi$_4$ & $1/4$ & 5.21 & 5.25 \\
        Co$_4$MnFeSi$_2$     & $1/2$ & 5.44 & 5.50 \\
        Co$_8$MnFe$_3$Si$_4$ & $3/4$ & 5.55 & 5.75 \\
        Co$_2$FeSi           & 1     & 5.56 & 6.00 \\
        \noalign{\smallskip}\hline
    \end{tabular}
    \label{tab1}
\end{table}

The results found from the LDA$+U$ calculations for the magnetic
moments are summarized in Tab.:~\ref{tab1} and compared to pure GGA
calculations without the inclusion of the $U$ type correlation. It is
evident from Tab.:~\ref{tab1} that the GGA derived values do not follow
the Slater-Pauling curve (with the exception of Co$_2$MnSi), whereas
the values from the LDA$+U$ follow the curve closely. These results
again indicate the loss of the minority gap - and thus the loss of
half-metallicity - if the on-site correlation is not included.

\subsection{Structural Properties}
\label{sec:SP}

Structural characterization has been performed with X-ray diffraction
(XRD) of powders as the standard method. Due to the small differences
in the scattering factors between the 3$d$ metals Mn, Fe and Co,
structural information, other than a simple confirmation of a single
cubic phase can only be gained by measuring the comparatively small
(5~{\%} of the (220)-peak) (111) and (200) superstructure peaks that
are typical for the face centered cubic (fcc) lattice. The simulated
powder diffraction pattern of Co$_2$MnSi shows the decisive (111) and
(200) peaks for the defect-free structure. Both of these super-lattice
peaks vanish for a random occupation of all lattice sites (4a, 4b, and
8c) resulting in the $A2$ structure. In the case of random occupation
of 4a and 4b sites by Mn, Fe, and Si, only the (200) super-lattice peak
of the $B2$ structure type would be seen, while the (111) peak would
vanish.

Such types of disorder would close the gap in the minority DOS so that
the material would no longer be a half-metallic ferromagnet
\cite{FKW05,WFK06b}. However, the magnetic moments may still follow a
Slater-Pauling like behavior. The half-metallic character is also
destroyed when only one of the Co atoms is exchanged by Mn/Fe ($X$
structure \cite{BPl71} with symmetry $F\:\overline{4}3m$). This type of
disorder shows up as a (111) super-lattice peak with higher intensity
than the (200) peak.

\begin{figure}
\centering
\includegraphics[width=7.5cm]{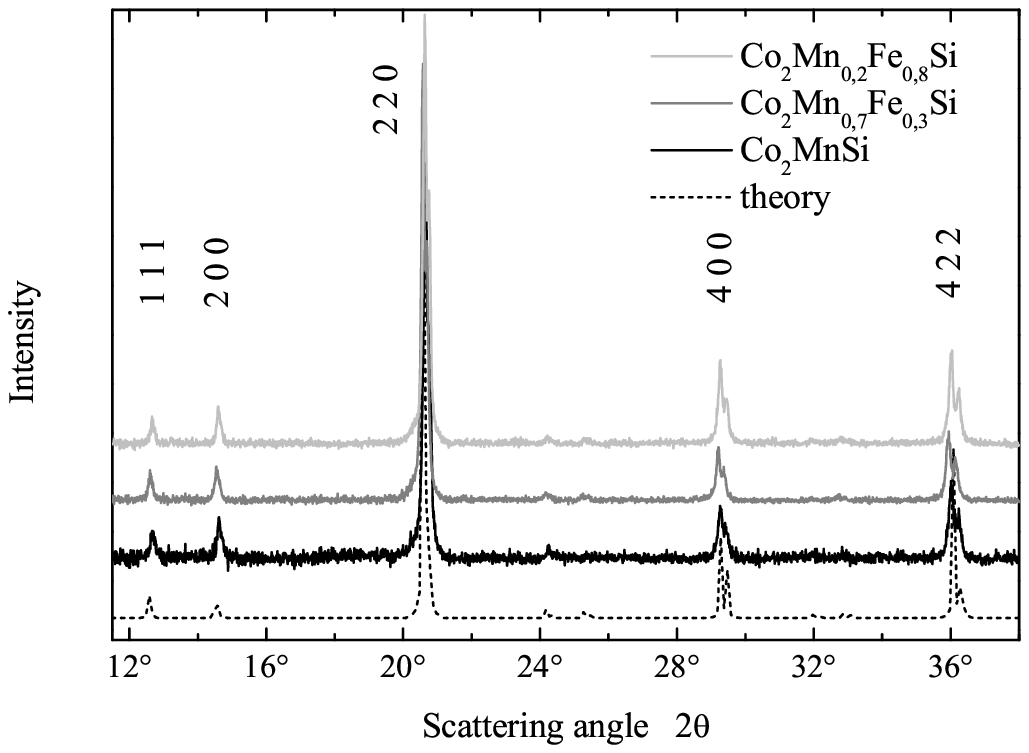}
\caption{XRD spectra for Co$_2$MnSi and Co$_2$Mn$_{0.8}$Fe$_{0.2}$Si. \\
         The spectra were excited by Mo K$_\alpha$ radiation.}
\label{fig3}
\end{figure}

As expected for the defect-free structure, the experimental data show
both the (111) and (200) peaks with equal intensity for all Fe
concentrations indicating the presence of a long range fcc structure
for all samples (see Fig.:~\ref{fig3}). Within the uncertainty of the
experiment, the lattice parameter of 5.64~{\AA} remains nearly
independent of the Fe concentration .

Because the scattering factors of all three transition metals are very
similar, X-ray diffraction cannot easily discern a disorder when the Mn
and Fe are partially exchanged with Co atoms on both 8a positions
($DO_3$ like disorder). Because both have the same $F\:m\overline{3}m$
symmetry, this leads to nearly identical diffraction patterns when
going from the $L2_1$ to the $DO_3$ structure. As will be shown in the
next section, this type of disorder can be ruled out by means of
M{\"o\ss}bauer spectroscopy.

\subsection{Magneto-structural Properties}
\label{sec:MSP}

\begin{figure}
\centering
\includegraphics[width=7.5cm]{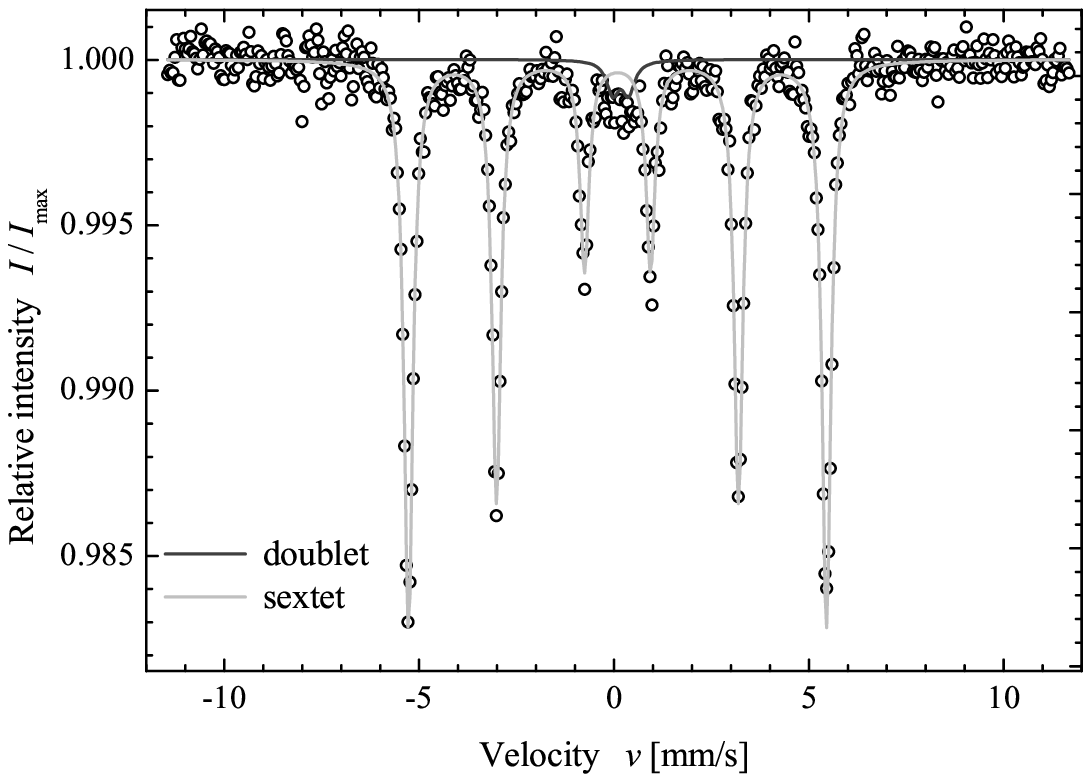}
\caption{$^{57}$Fe M{\"o\ss}bauer spectrum of Co$_2$Mn$_{0.5}$Fe$_{0.5}$Si. \\
         The spectrum was taken at 290K and excited by a $^{57}$Co(Rh) source.
         Solid lines are results of a fit to determine the sextet and doublet
         contributions and to evaluate the hyperfine field.}
\label{fig4}
\end{figure}

$^{57}$Fe M{\"o\ss}bauer spectroscopy was performed to investigate the
magneto-structural properties. The transmission spectrum of
Co$_2$Mn$_{0.5}$Fe$_{0.5}$Si is shown in Fig.:~\ref{fig4}. Starting
from 10~{\%}, the spectra for the complete range of Fe concentration
$x$ are all similar and therefore not shown here. The observed
sextet-like pattern is typical for a magnetically ordered system. The
pattern is typical for the cubic symmetry and no asymmetric shift of
the lines from a non-cubic quadrupole interaction is observed.

The spectrum shown in Fig.:~\ref{fig4} for 50~{\%} Fe is dominated by
an intense sextet. With Fe occupying the 4a sites with cubic symmetry
($O_h$), this sextet indicates on the high order of the sample. In
addition to the sextet, a much weaker line at the center of the
spectrum is visible. Depending upon the composition it can be explained
as a singlet or doublet. Its contribution to the overall intensity of
the spectrum is approximately 3.5~{\%} at $x=0.5$. The origin of the
singlet or doublet may be caused by anti-site disorder leading to a
small fraction of paramagnetic Fe atoms. The splitting of the
paramagnetic line into a doublet is due to dynamic effects and, among
others reasons, usually depends upon the size of the powder grains. The
slight disorder arises most likely from surface regions of the sample
that are destroyed when the sample is crushed to powder. A partial
contamination of the relatively large surface area of the powder with
oxygen can also not been excluded. The relative contribution of the
doublet decreases exponentially from 9~{\%} in low Fe-substituted
Co$_2$Mn$_{0.9}$Fe$_{0.1}$Si to 1.8~{\%} in pure Co$_2$FeSi. As was
also suggested by the photo absorption and ESCA measurements, this may
indicate a larger probability for oxidation in the Mn rich part of the
series.

The line width of the sextet is approximately $(0.14\pm0.01)$~mm/s
(corresponding to 6.7~neV) on average over the complete series of
compositions. A pronouncedly higher line width of $\approx0.19$mm/s was
found for the alloy with $x=0.7$. This indicates a higher disorder in
that sample and may also explain the slightly higher magnetic moment
(compare Fig.:~\ref{fig8} in Sec.:~\ref{sec:MP}). The isomer shift
increases linearly from 0.075~mm/s (3.6~neV) to 0.129~mm/s (6.2~neV)
with increasing $x$, indicating the change in the environment of the Fe
atoms, that appears here in the second nearest neighbor shell where the
next Fe or Mn atoms are located. Despite this increase, the values
suggest a Fe$^{3+}$-like character of the iron atoms in
Co$_2$Mn$_{1-x}$Fe$_{x}$Si. The increase points on a slight decrease of
the valence electron concentration close to the iron atoms. Table
\ref{tab2} summarizes the values for the isomer shift and the hyperfine
field ($H_{hff}$) as function of the iron concentration.

\begin{table}
\centering \caption{M{\"o\ss}bauer data for iron in Co$_2$Mn$_{1-x}$Fe$_x$Si. \\
                    Given are the measured and calculated values of the hyperfine field
                    ($H_{hff}$) and the measured isomer shift ($IS$)
                    for increasing Fe concentration $x$.}

     \begin{tabular}{l|c|c|c}
        $x$          & $H_{hff}$  [10$^6$~A/m] &              & $IS$ [neV] \\
                     & Experiment              & Calculation  &            \\
        \noalign{\smallskip}\hline\noalign{\smallskip}
        0.1          &  25.937    &         & 3.61   \\
        0.2          &  26.214    &         & 3.77   \\
        0.25         &  -         & 26.534  &        \\
        0.4          &  26.412    &         & 4.18   \\
        0.5          &  26.466    & 26.629  & 4.34   \\
        0.6          &  26.417    &         & 4.75   \\
        0.7          &  26.259    &         & 4.90   \\
        0.75         &  -         & 26.962  &        \\
        0.8          &  25.915    &         & 5.33   \\
        0.9          &  25.480    &         & 5.77   \\
        1            &  24.997    & 27.013  & 6.21   \\
        \noalign{\smallskip}\hline
    \end{tabular}
    \label{tab2}
\end{table}

The hyperfine field at the Fe sites amounts to $26.5\times10^6$~A/m in
Co$_2$Mn$_{0.5}$Fe$_{0.5}$Si. This is the maximum value observed in the
complete series with varying Fe concentration $x$. Overall, the
hyperfine field varies non-linearly from $25.9\times10^6$~A/m at
$x=0.1$ to $25\times10^6$~A/m at $x=1$ (see also: \footnote{Note that
the M{\"o\ss}bauer data reported in Ref.\cite{WFK05} were taken at
lower temperature (85K).}). For low iron concentration, it increases
with $x$ and decreases from $x=0.5$ to $x=1$. It should be noted that
the M{\"o\ss}bauer spectra taken at 85K from Co$_2$FeSi exhibited a
considerably higher value ($26.3\times10^6$~A/m) without additional
singlet or doublet contributions. Therefore, thermally activated
fluctuations or disorder can not be excluded here. The values of the
hyperfine field at the Fe atoms are comparable to those found by
Niculescu {\it et al} \cite{NBR77,NBH79} using spin-echo nuclear
magnetic resonance (NMR). For Co$_{3-x}$Fe$_x$Si, these authors
reported approximately $26.9\times10^6$~A/m for iron on 4a sites. The
values for partial occupancy of 8c sites, expected from NMR
\cite{NBH79} ($\approx16\times10^6$~A/m) for Co$_2$FeSi are
considerably smaller. This is in agreement with calculations for Co and
Fe on interchanged sites ($\approx17\times10^6$~A/m). Therefore, a
$DO_3$ type disorder can be excluded. The calculated hyperfine fields
are, however, nearly independent of the Fe concentration. They decrease
linearly with $x$ by $-0.7\times10^6$~A/m from $27.01\times10^6$~A/m
for Co$_2$FeSi. A maximum in the $H_{hff}(x)$ dependence at $x=1/2$
could not be verified. It was found neither for the ordered compounds
using the FLAPW method with the LDA$+U$ (see Tab.: \ref{tab2}), nor for
random alloys calculated using a KKR-CPA scheme in the GGA
approximation \footnote{Note that these calculations need at present a
rather unphysical enlargement of the lattice parameter in order to
explain the magnetic moments and position of the minority gap
correctly.}.

\begin{figure}
\centering
\includegraphics[width=7.5cm]{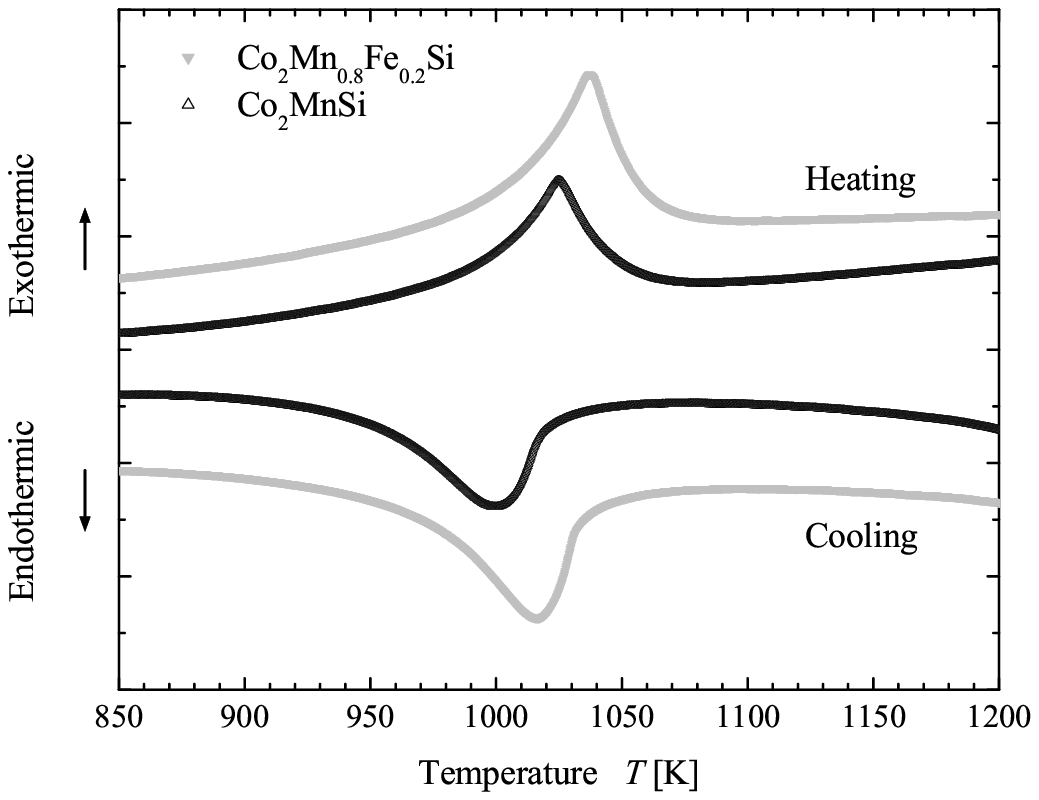}
\caption{DSC results for Co$_2$MnSi and
         Co$_2$Mn$_{0.8}$Fe$_{0.2}$Si.}
\label{fig5}
\end{figure}

\begin{figure}
\centering
\includegraphics[width=7.5cm]{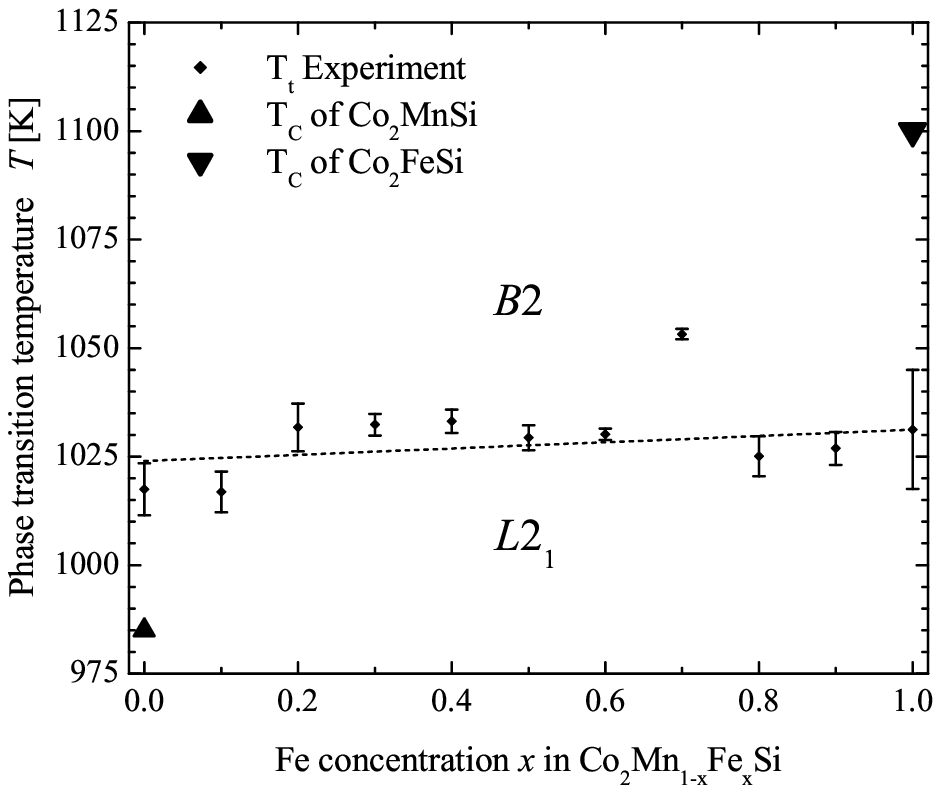}
\caption{Phase transitions in Co$_2$Mn$_{1-x}$Fe$_{x}$Si.\\
         The straight line is the result of a linear fit of $T_t(x)$ as function
         of the Fe concentration. The given Curie temperatures are taken from
         Refs. \cite{Web71} and \cite{WFK05} for Co$_2$MnSi and Co$_2$FeSi,
         respectively.}
\label{fig6}
\end{figure}

Differential scanning calorimetry was used to find the high temperature
phase transitions in the substitutional series. Figure \ref{fig5} shows
a typical result from DSC, which was used to investigate the expected
phase transitions. The figure displays the change of the DSC signal as
a function of the temperature using nominal heating and cooling rates
of 20~K/min. A strong signal arising from a phase transition is easily
detected at about 1000~K to 1050~K during both heating and cooling. The
shift of the maxima is mainly due to an intrinsic hysteresis effect of
the method and depends on the temperature rates and the actual amount
of material. The length of the error bars in Fig.:~\ref{fig6}
corresponds to this hysteresis.

A series of DSC measurements was performed with different scanning
rates for heating and cooling, but it was not possible to distinguish
the magnetic transition temperature because it was too close to the
structural transition temperature of the $L2_1$ to the $B2$ phase
$T_t^{B2 \leftrightarrow L2_1}$. To overcome the problem of nearby
phase transitions of different kind, it would be necessary to obtain
high temperature magnetization curves such as those that Kobayashi et
al. \cite{KUF05} obtained in their examination of the series
Co$_2$Cr$_{1-x}$Fe$_x$Ga, or the difference in the transition
temperatures should be more than 100~K.

Figure \ref{fig6} displays the temperature dependence of the $L2_1
\leftrightarrow B2$ phase transition. It is nearly constant, increasing
slightly, as the Fe content increases, from $\approx1024$~K for
Co$_2$MnSi to $\approx1031$~K for Co$_2$FeSi. It is seen that the Curie
temperatures of the end members of the substitutional series are
slightly below or above the structural phase transition for the
compound containing Mn or Fe, respectively. The Curie temperature is
expected to increase with increasing $x$ from $T_C(0)=985$~K
\cite{Web71} to $T_C(1)=1100$~K \cite{WFK05}, while in
Co$_2$Mn$_{1-x}$Fe$_x$Si $T_t^{B2 \leftrightarrow L2_1}$ hardly varies
with increasing $x$, being  around 1027~K for $x=0.5$. It is therefore
impossible here to unambiguously determine $T_C$ by using the DSC
technique, because of the relative weakness of the magnetic transition
compared to the structural transition and the overlap of those two
transitions in the DSC spectra. It is interesting to note that the
Curie temperature of the compounds with high Fe concentration appears
to be above the order-disorder phase transition.

\subsection{Magnetic Properties}
\label{sec:MP}

The Co$_2$-based Heusler alloys that are half-metallic ferromagnets
show a Slater-Pauling like behavior for the magnetization (see
Sec.:~\ref{sec:EMS}). The saturation magnetization scales with the
number of valence electrons \cite{FKW06} and the magnetic moment per
unit cell is given by Eq.:~\ref{eq1}. A magnetic moment of:

\begin{equation}
   m(x) = (5 + x) \mu_B
\label{eq2}
\end{equation}

is expected for Co$_2$Mn$_{1-x}$Fe$_x$Si.

\begin{figure}
\centering
\includegraphics[width=7.5cm]{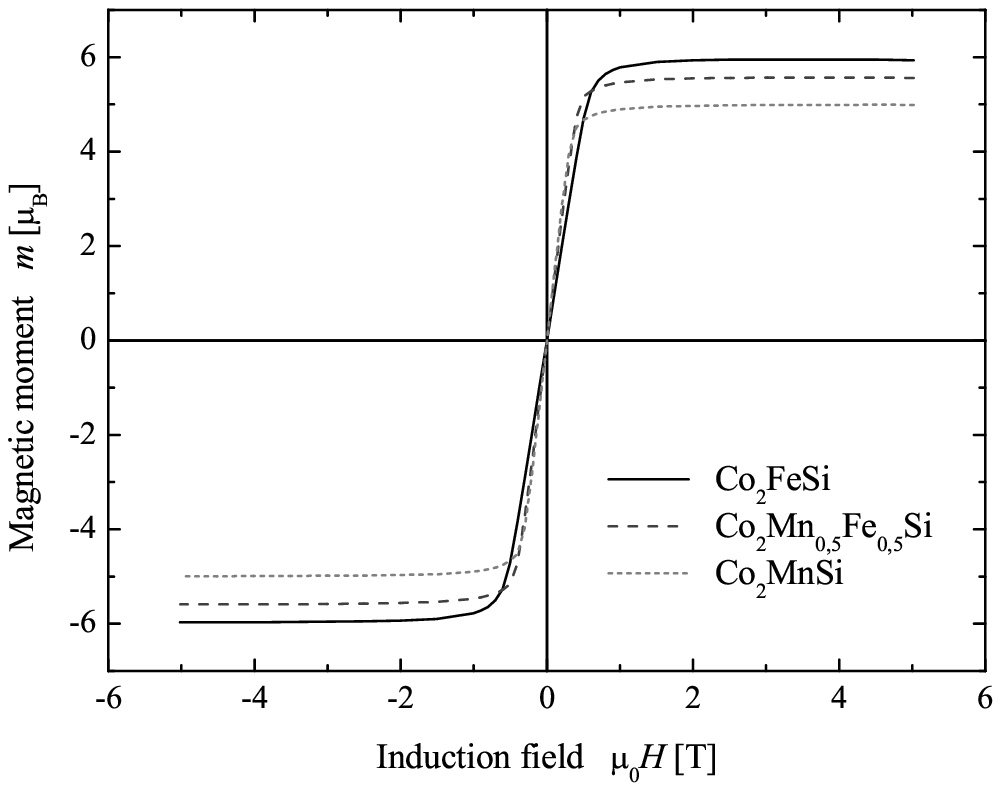}
\caption{Magnetization of Co$_2$Mn$_{1-x}$Fe$_{x}$Si.\newline
         Displayed are the hysteresis curves for $x=0,0.5,1$
         taken at $T=5$~K.}
\label{fig7}
\end{figure}

Low temperature magnetometry was performed by means of the SQUID to
check the calculated saturation moment. Selected results are shown in
Fig.:~\ref{fig7}. The increase of the saturation moment with the iron
concentration is clearly visible. In addition, it is found that all
Co$_2$Mn$_{1-x}$Fe$_x$Si samples are soft magnetic with a small
remanence and a small coercive field. Results for the element-specific
magnetic moments from X-ray magnetic circular dichroism are reported
elsewhere \cite{KEB06}.

\begin{figure}
\centering
\includegraphics[width=7.5cm]{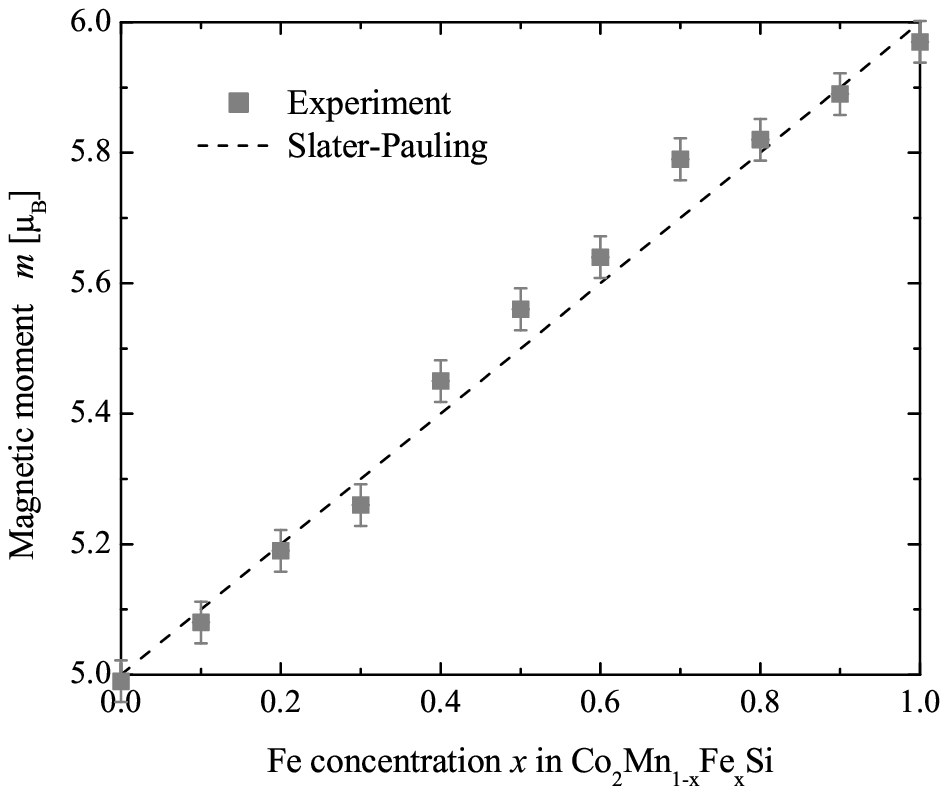}
\caption{Concentration dependence of the magnetic moment in
         Co$_2$Mn$_{1-x}$Fe$_{x}$Si.\newline
         All measurements were performed at $T=5$~K.}
\label{fig8}
\end{figure}

The total magnetic moments, measured at 5~K and in saturation, are
$(4.97 \pm 0.05)\mu_B$ and $(5.97 \pm 0.05)\mu_B$ for the pure
compounds Co$_2$MnSi and Co$_2$FeSi, respectively. The latter is in
perfect agreement with the recent investigation reported in
Refs.\cite{WFK05,WFK06a}. Figure \ref{fig8} displays the dependence of
the saturation moment as a function of the Fe concentration $x$. The
series shows a nearly linear increase of $m$ with increasing Fe
concentration that closely fits the values expected from a
Slater-Pauling like behavior.

Comparing the experimental results to the theoretical values as given
in Tab.:~\ref{tab1}, it is evident that they closely agree with those
from the LDA$+U$ calculations. The agreement of the GGA result for
Co$_2$MnSi may thus be seen as due to chance. The comparison also
substantiates the use of correlation energies of about 0.135~Ry, as
these can be used to predict the magnetic moment correctly over the
entire range of Fe concentration $x$.

\subsection{Electronic Properties}
\label{sec:EP}

\begin{figure*}
\centering
\includegraphics[width=12.5cm]{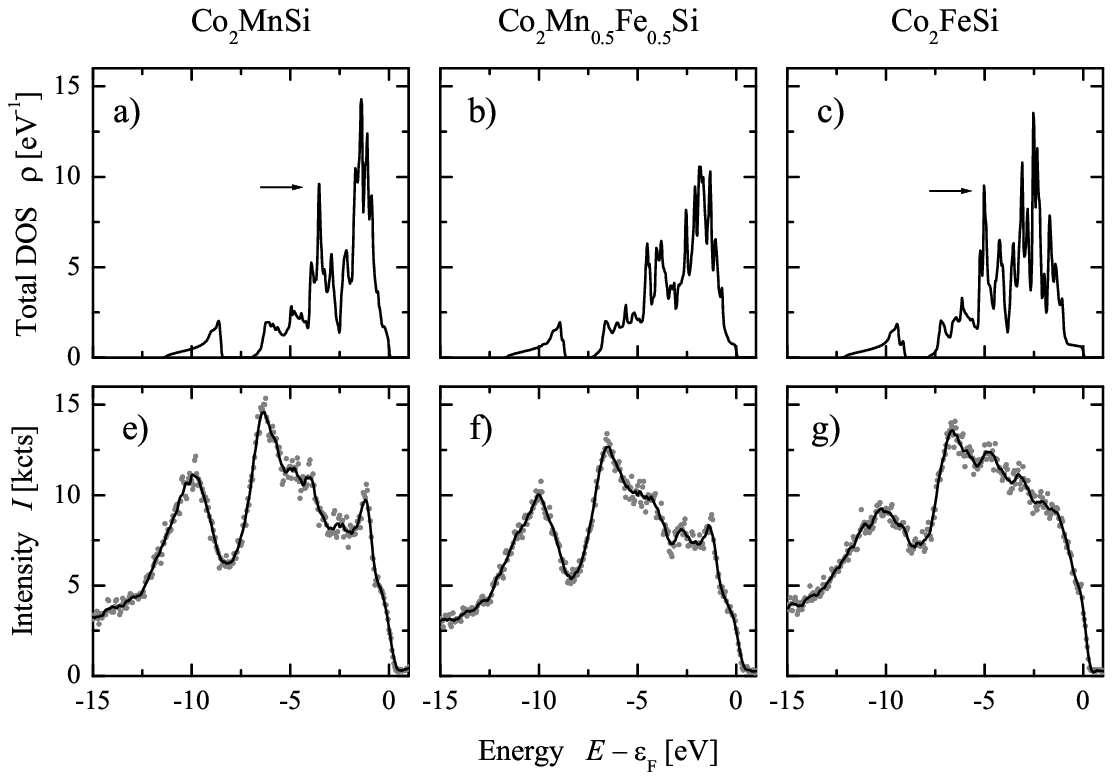}
\caption{Valence density of Co$_2$Mn$_{1-x}$Fe$_{x}$Si ($x=0, 1/2, 1$).\newline
         The calculated total density of states (a, b, c) has been convoluted by a Fermi-Dirac
         distribution using $T = 20$~K.
         The photoelectron spectra (e, f, g) have been excited by $h\nu=7.939$~keV.}
\label{fig9}
\end{figure*}

The results from high energy photo emission are shown in Figure
\ref{fig9} and compared to the total density of states calculated for
Co$_2$Mn$_{1-x}$Fe$_{x}$Si with $x=0, 1/2, 1$). The calculated total
DOS is the sum of the spin resolved majority and minority DOS shown in
Figures \ref{fig1} \& \ref{fig2} and convoluted with a Fermi-Dirac
distribution using $T=20$~K.

The spectra of all three compounds reveal clearly the low lying
$s$-states at about -11~eV to -9~eV below the Fermi energy, in well
agreement to the calculated DOS. These low lying bands are separated
from the high lying $d$-states by the Heusler-typical hybridization gap
being clearly resolved in the spectra as well as the calculated DOS.
The size of this gap amounts typically to $\Delta E\approx 2$~eV in Si
containing compounds.

Obviously, the emission from the low lying $s$-states is pronouncedly
enhanced compared to the emission from the $d$-states. This can be
explained by a different behavior of the cross sections of the $s$,
$p$, and $d$ states with increasing kinetic energy as was recently
demonstrated by Panaccione {\it et al} for the case of the silver
valence band \cite{PCC05}. In particular, the cross section for
$d$-states decreases faster with increasing photon energy than the one
of the $s$-states. This behavior influences also the onset of the
$d$-bands at about -7~eV. Just at the bottom of those $d$-bands, they
are hybridized with $s, p$-like states, leading to a high intensity in
this energy region.

The structure of the spectra in the range of the $d$ states agrees with
the structures observed in the total DOS. However, one has to account
for lifetime broadening and the experimental resolution if comparing
that energy range. The lowest flat band of the majority band structure
(see Fig.:\ref{fig1}), accompanying the localized moment at the Y
sites, results in a sharp peak in the DOS at about -3.5~eV and -5~eV
for Mn and Fe, respectively (marked by arrows in Fig.\ref{fig9} a) and
c). These peaks are shifted away from $\epsilon_F$ by the
electron-electron correlation in the LDA$+U$ calculation and would
appear without $U$ closer to the Fermi energy. Their energetic position
corresponds to structures revealed in the measured spectra, thus they
are a good proof for the use of the LDA$+U$ scheme.

Most interesting is the behavior of the calculated DOS and the measured
spectra close to $\epsilon_F$ as this might give an advice about the
gap in the minority states. The majority band structure contributes
only few states to the density at $\epsilon_F$ emerging from strongly
dispersing bands. This region of low density is enclosed by a high
density of states arising from flat bands at the upper and lower limits
of the minority band gap. The onset of the minority valence band is
clearly seen in the total DOS as well as the low majority density at
the Fermi energy. The same behavior is observed in the measured valence
band spectra. From the spectra, it can be estimated that the Fermi
energy is in all three cases about 0.5~eV above the minority valence
band. This gives strong evidence that all compounds of the
Co$_2$Mn$_{1-x}$Fe$_x$Si series exhibit really half-metallic
ferromagnetism.

The values for $U_{eff}$ used in section \ref{sec:EMS} are the
borderline cases for the half-metallic ferromagnetism over the complete
series Co$_2$Mn$_{1-x}$Fe$_x$Si. They were used being independent of
the Fe concentration, what was suggested for Co from the constrained
LDA calculations. However, the valence band spectra shown in
Figure~\ref{fig9} indicate that the Fermi energy of both end members
may fall inside of the minority gap rather than being located at the
edges of the minority gap. This situation may be simulated by a
variation of $U$. A comparison to the $U$-dependence of the minority
gap shown in Ref.:~\cite{KFF06} suggests smaller effective Coulomb
exchange parameters for the Mn rich part and larger ones for the Fe
rich part of the series. This might also explain the non-linearity
reported in section~\ref{sec:MP} for the hyperfine field. A variation
of those parameters for all contributing $3d$ constituents in the
calculations was omitted here because it would not bring more insight
into the nature of the problem, at present.

Overall, the measured photoelectron spectra agree well with the
calculated density of states and principally verify the use of the
LDA$+U$ scheme. In particular, the shape of the spectra close to
$\epsilon_F$ can be explained by the occurrence of a gap in the
minority states and thus points indirectly on the half-metallic state
of all three compounds investigated here by photo emission. For clarity
about the gap, spin resolved photo emission spectroscopy at high
energies would be highly desirable. However, this will make another
step of improvement of the instrumentation necessary, for both photon
sources as well as electron energy and spin analyzers, as the spin
detection will need a factor of at least three to four orders of
magnitude more intensity for a single energy channel at the same
resolution as used here for the intensity spectra.

\section{Summary and Conclusion}

The substitutional series of the quaternary Heusler compound
Co$_2$Mn$_{1-x}$Fe$_x$Si was synthesized and investigated both
experimentally and theoretically.

The results found from the LDA$+U$ calculations for the magnetic
moments $m(x)$ closely follow the Slater-Pauling curve. The shift of
the minority gap with respect to the Fermi energy, from the top of the
minority valence band for Co$_2$MnSi to the bottom of the minority
conduction band for Co$_2$FeSi, makes both systems rather unstable with
respect to their electronic and magnetic properties. The calculated
band structures suggest that the most stable compound in a
half-metallic state will occur at an intermediate Fe concentration.
These theoretical findings are supported by the experiments.

All samples of the substitutional series exhibit an $L2_1$ order that
is independent of the Fe concentration $x$. The observed structural
order-disorder phase transition from $L2_1 \leftrightarrow B2$ is
nearly independent of $x$ and occurs at about 1030~K. M{\"o\ss}bauer
measurements show only a negligible paramagnetic contribution
confirming the high degree of order over the whole substitutional
series. In agreement with the expectation from the Slater-Pauling
curve, the magnetic moment increases linearly with $x$ from 5~$\mu_B$
to 6~$\mu_B$. True bulk sensitive, high energy photo emission bearded
out the inclusion of electron-electron correlation in the calculation
of the electronic structure and gave an indirect advise on the gap in
the minority states. Both, valence band spectra and hyperfine fields
indicate an increase of the effective Coulomb exchange parameters with
increasing Fe concentration.

From both the experimental and computational results it is concluded
that a compound with an intermediate Fe concentration of about 50\%
should be most stable and best suited for spintronic applications.

\bigskip
\begin{acknowledgments}

The authors are grateful for the support by G.~Sch{\"o}nhense and thank
J.~Thoenies and A.~Lotz for help with the calculation of $U_{eff}$ as
well as V.~Jung, V.~Ksenofontov, S.~Reimann (Mainz) for performing the
M{\"o\ss}bauer experiments, and Y.~Takeda (Spring-8, Japan) for help
during the beamtime. The synchrotron radiation experiments were
performed at the beamline BL47XU of SPring-8 with the approval of the
Japan Synchrotron Radiation Research Institute (JASRI) (proposal no.
2006A1476). This work was partially supported by a Grant-in-Aid for
Scientific Research (A) (No.15206006), and also partially supported by
a Nanotechnology Support Project, of The Ministry of Education,
Culture, Sports, Science and Technology of Japan. Financial support by
the Deutsch Forschungs Gemeinschaft (projects TP1 and TP7 in research
group FG 559) is gratefully acknowledged.

\end{acknowledgments}

\bibliography{balke_Co2MnFeSi}

\end{document}